\documentclass[aps,11pt,final,notitlepage,oneside,onecolumn,nobibnotes,nofootinbib,preprintnumbers,superscriptaddress,showpacs,centertags,showkeys,amsmath,amssymb]{revtex4}

\usepackage{graphicx} 
\usepackage{dcolumn} 

\def\phi{\varphi}

\begin {document}

\title {Study of an Abelinization Transition in SU(2) Gluodynamics at Finite Temperature}
\author{J. Boh\' a\v cik}
\email{bohacik@savba.sk} \affiliation{Institute of Physics, Slovak
Academy of Sciences, D\' ubravsk\' a cesta 9, 811 45 Bratislava,
Slovakia.}
\author{P. Pre\v snajder}
\email{presnajder@fmph.uniba.sk} \affiliation{Department of
Theoretical Physics and Physics Education, Faculty of Mathematics,
Physics and Informatics, Comenius University, Mlynsk\' a dolina F2,
842 48 Bratislava, Slovakia.}

\date{\today}

\begin{abstract}
We discuss the problem of an effective descriptions of the
phase transition phenomena in the pure gluodynamics in $SU(2)$ symmetric QCD. We
choose the method of calculation following
 the conjecture that the infrared sector of the theory
possesses the same confinement characteristic as the full theory. We
show, that analytic descriptions of this phenomena is beyond the
Gaussian method of evaluations of functional integrals. We propose a
non-perturbative evaluation of functional integral, meanwhile for
two dimensional Wiener integral for $x^4$ theory.
\end{abstract}

\pacs{11.15.-q, 11.15.Tk, 11.25.Db}
\keywords{path integral,
gluodynamics, phase transitions}

\maketitle

\section{Introduction}

The evaluations of the measurable quantities in  quantum field
theory (QFT) are well -- defined theoretically, but in real
calculations appear many obstacles connected with the non-ability to
realize the calculations corresponding to the theoretical
definitions. The calculations in QFT are usually performed
analytically using perturbative calculus, numerically by lattice
gauge theory (LGT), or by creating models based on the results of
previous methods of calculations. The perturbative methods give
analytical results, but various problems in QFT are non-perturbative
phenomena, therefore they are applicable for the narrow band of
problems only. LGT, the best method of evaluations in this time,
suffers from finiteness of the lattice, problems of definitions of
the mathematical objects on the lattice and the continuum limit
problems.  We adopt the method of analytical calculations in the
continuum to study of the transitions in dense matter. After short
introduction into the problem in the next section we propose in the
section 3 the non-perturbative solution of the problem of the
evaluation of the functional integral with fourth order term in the
action. We found the result in the form of an asymptotical series,
and a method of solution of the problem is described in papers
\cite{paper1} and \cite{paper2}.

\section{Effective description of the confinement in continuum}

The Euclidean finite temperature theory is defined by the functional
integral for partition function:
$$Z(\beta)= N\int\; [D\mathcal{A}^{a}_{\mu}]\; \exp{(-S_E)}$$
where the Euclidean action is defined by:
$$S_E = \frac{1}{4}\int_0^{\beta}\; d \tau \int d^3x\; \mathcal{F}^a_{\mu \nu}\mathcal{F}^a_{\mu \nu}$$
The Euclidean color field strength is defined by:
$$\mathcal{F}^a_{\mu \nu} = \partial_{\mu} \mathcal{A}_{\nu}^a - \partial_{\nu} \mathcal{A}_{\mu}^a
- g\varepsilon^{abc}\mathcal{A}_{\mu}^b \mathcal{A}_{\nu}^c$$ We
require the periodicity of color potentials in the direction of the
imaginary time variable:
$$\mathcal{A}_{\mu}^a(\tau+\beta,x) = \mathcal{A}_{\mu}^a(\tau,x)\  .$$
The theory with corresponding partition function looks like a
partition for the finite temperature theory in statistical physics.
This "isomorphism" allows us to investigate the QCD problem by
methods of statistical physics.

We have concerned on the phase transitions to the quark gluon plasma
at high densities of the matter, the one of the challenging forecast
of the QCD. We investigate some problems of the study of this phase
transition, usually known as confinement/deconfinement phase
transition, in model independent method described bellow. Such
description of the confinement problem rely on the conjecture that
the infrared sector of the finite temperature theory possesses the
same phase structure as the complete theory. This is the assertion
of Appelquist - Carrazone decoupling theorem \cite{apca}, where it
was proven that integrating off massive modes of the theory one
obtains terms in the potential controlling dynamics (in this case
the infrared variables) of the theory. By help of this theorem we
reduce the infinite number of the variables of the original theory
to the finite number in the infrared sector effectively describing
the phenomena studied.

To evaluate the integration over the massive modes we proceed as
follows:

1) We separate the field variable to the sum of the "classical" part
and "quantum fluctuation":
$$\mathcal{A}_{\mu}^a(t,x) = A_{\mu}^a(t,x) + a_{\mu}^a(t,x).$$
We use the static gauge for evaluations of the functional integrals
over massive modes, The saddle points of the action in this gauge
are fixed as:
$$A_i^a = 0,\ i,a=1, 2, 3,\ A_0^1 =A_0^2 =0, \ A_0^3 = const. $$
We choose finite $A_0^3$ in the color space for the
simplicity.

2) The periodicity of the color potentials in the imaginary time
variable offers a chance to use Fourier transformation of the
potentials in this variable. This operation automatically separates
the Fourier transforms of the potentials to the infrared, zero
frequency modes and massive modes, where nonzero fourier frequency
modes ($\equiv$ Matsubara frequencies) appear as mass terms.
Integrating over this "massive modes" by Gaussian method  we obtain
the contribution to the effective potential (in the scope of this
article named $V_{eff}=V_{eff}(A^3_0, b^3_0),\ b^a_i, \ a=i=1, 2,
3$) and contributions of the $A^3_0$ dependent mass terms for the
perpendicular quantum fluctuation degrees of freedom $(b^1_i, b^2_i,
i=1, 2, 3)$ in the infrared sector of the theory. This process is
known as dimensional reduction, described by Appelquist and Pisarski
\cite{appi}, offers possible calculation scheme for finite
temperature field theories. The zero modes action in quantum
fluctuations fields read:
\begin {eqnarray}
S_{zero}=\beta \int d^3x \{\frac{1}{4}G^a_{ij}G^a_{ij} +
\frac{1}{2}(gA^3_0)^2[(b^1_i)^2 + (b^2_i)^2] +\frac{1}{2}(\partial_i
b^3_0)^2 + V_{eff}(A^3_0, b^3_0), b^a_i)\} \label{eq1}
\end {eqnarray}
with field strength
$$G^a_{ij} = \partial_i b_j^a - \partial_j b_i^a - g\varepsilon^{abc}b_i^b b_j^c$$

We have found the two different effective systems, depending on the
value of the static color potential $A^3_0$.

1. Let $A^3_0 = 0.$ In this case all quadratic terms in zero
frequency field variables in Eq.(\ref{eq1}) disappears. The
effective infrared system depend on variables $b^a_i, b^3_0,\;
a,i=1, 2, 3.$ Color $SU(2)$ symmetry of the system is maintained.
This system correspond to interaction of the effective Higgs field,
represented by chromoelectric potential $b^3_0$, with chromomagnetic
fields represented by chromomagnetic potentials $b^a_i$.

2. Let $A^3_0 \gg 0,$ at least as lowest Matsubara frequency. In
this case  the modes $b^1_i, b^2_i$ in Eq. (\ref{eq1})  must be
treated as massive, and we integrate off these modes from the zero
mode system. Then the infrared system depend on variables $b^3_i,
b^3_0,\; i=1, 2, 3$ and the remanent color symmetry group is $Z(2).$
It is the well-known abelian projection observed in the lattice
calculations also. In earlier continuum theory studies \cite{bo1},
\cite{bopr}, \cite{bobo}, we find in this sector the
confinement/deconfinement transition with correct behavior of the
potential between heavy quarks calculated as the corellators of
Polyakov's loops.

But what happens when $A^3_0$ grows from zero to a finite value and
infrared system undergo transition $SU(2)\rightarrow Z(2)$? This
phenomenon we named as "abelinization transition" \cite{dig}. For
its studies the use the Gaussian method for evaluations of
functional integrals is not sufficient. Preliminary investigations
indicate that this transition could be non-continuous, and
consequently the lattice formalism may be not powerful also. We
conclude, that one needs a formalism for the evaluations of the
functional integrals with a higher power terms in the potential.
This formalism enables to eliminate the mass term, even in this
limit the functional integral is well defined.

As a first attempt we applied the method to evaluate (semi)
analytically Wiener functional integral with $x^4$ term in the
action. The short description of this method is reported in the next
section.

\section{Evaluation of $\varphi^4$ Wiener functional integral}

The simplest non-gaussian functional integral is Wiener functional
integral with $x^4$ term in the action. In Euclidean sector of the
theory we have to evaluate the continuum Wiener functional integral:
\begin {equation}
\mathcal{Z} = \int [\mathcal{D}\varphi(x)]\exp (-\mathcal{S})\ ,
\label{fi}
\end{equation}

In this case the action possesses the fourth order term:
\begin {equation}
\mathcal{S} =\int \limits _0^\beta d\tau \left[c/2
\left(\frac{\partial\varphi(\tau)}{\partial\tau}\right)^2+b\varphi(\tau)^2
+a\varphi(\tau)^4\right]\ . \nonumber
\end {equation}

The continuum Wiener  functional integral is defined by a formal
limit:
$$\mathcal{Z} = \lim_{N\rightarrow \infty}\;\mathcal{Z}_{N}\ .$$

Here the finite dimensional integral $\mathcal{Z}_{N}$ is defined by
the time -- slicing method:

\begin {eqnarray}
\mathcal{Z}_{N}=\int\limits _{-\infty}^{+\infty} \prod \limits
_{i=1}^N\left(\frac{d\varphi_i}{\sqrt{\frac{2\pi\triangle}{c}}}\right)
\exp\left\{-\sum\limits _{i=1}^N
\triangle\left[c/2
\left(\frac{\varphi_i-\varphi_{i-1}}{\triangle}\right)^2
+b\varphi_i^2+a\varphi_i^4\right]\right\}
\label{ndim}
\end {eqnarray}
\noindent where $\triangle=\beta/N$, and $a, b, c$ are the
parameters of the model. The quantity $Z_N$ represents the
unconditional propagation from $\varphi=\varphi_0 =0$ to any
$\varphi=\varphi_N$ (Eq. (\ref{ndim}) contains an integration over
$\varphi_N$).

\subsection{One dimensional integral}

An important task is to calculate the one dimensional integral
\begin {equation}
I_1=\int\limits _{-\infty}^{+\infty}\;dx\;\exp\{-(A x^4+B x^2+C
x)\}\ ,
 \nonumber
\end {equation}
where $Re\: A>0$. The standard perturbative procedure rely on
Taylor's decomposition of $\exp(-Ax^4)$ term with consecutive
replacements of the integration and summation order. The integrals
can be calculated, but the sum is divergent.

Instead we propose the power expansion in $C$:
\begin {equation}
I_1=\sum\limits _{n=0} ^{\infty} \frac{(-C)^n} {n!}\int\limits
_{-\infty}^{+\infty}\;dx\;x^n\exp\{-(A x^4+B x^2)\}\ .
 \nonumber
\end {equation}

The integrals in the above relation can be expressed by the
parabolic cylinder functions $D_{-\nu-1/2}(z).$ Then, the integral
$I_1$ can be reds:
\begin {equation}
 I_1=\frac{\Gamma(1/2)}{\sqrt{B}}\sum\limits _{m=0} ^{\infty}
 \frac{\xi^m}{m!}\mathcal{D}_{-m-1/2}(z)\ ,\
\xi=\frac{C^2}{4B}\ ,\
 z=\frac{B}{\sqrt{2A}}\ ,
\label{lb1}
\end {equation}

\noindent
where we used the abbreviation:
$$
 \mathcal{D}_{-m-1/2}(z) =  z^{m+1/2}\;
e^{\frac{\scriptstyle z^2}{\scriptstyle 4}}\; D_{-m-1/2}(z)\
.\nonumber
$$
\noindent
 It was shown, that sum in the Eq. (\ref{lb1}) is
convergent and for finite values $z$ this sum converges uniformly in
$\xi$.

Applying this idea of integration to the $N$ dimensional integral
(\ref{ndim}) we proved \cite {paper1} the exact formula:
\begin {equation}
\mathcal{Z}_{N} = \left[\prod_{i=0}^{N}2(1+b\triangle^2/c)
\omega_i\right]^{-\frac{1}{2}} \; \mathcal{S}_{N} \nonumber
\end {equation}
with
\begin {equation}
\mathcal{S}_{N}= \sum\limits_{k_1,\cdots,k_{N-1}=0}^\infty \prod
\limits _{i=0}^N \; \left[
\frac{\left(\rho\right)^{2k_{i}}}{(2k_{i})!}
\Gamma(k_{i-1}+k_{i}+1/2)\; \sqrt{\omega_i}\;
\mathcal{D}_{-k_{i-1}-k_{i}-1/2}\;(z)\right], \nonumber
\end {equation}

\noindent
where $k_0 = k_N = 0,$ $\rho=(1+b\triangle^2/c)^{-1}$,
$z=c(1+b\triangle^2/c)/\sqrt{2a\triangle^3}$,
$\omega_i=1-A^2/\omega_{i-1}$, $\omega_0=1/2+A b\triangle^2/c$,
$A=\frac{1}{2(1+b\triangle^2/c)}.$ We see that

\noindent - $\rho$ is independent of the coupling constant.

\noindent - Only the argument $z$ of parabolic cylinder function is
coupling dependent.

To evaluate $S_N$, we must solve the problem how to sum up the
product of two parabolic cylinder functions. The parabolic cylinder
functions are related to the representation of the group of the
upper triangular matrices, so we implicitly expect a simplification
of their product. This problem is not solved completely yet. We
adopt less complex method of summation, namely we use the asymptotic
expansion of one of them, then, exchanging the order of summations
we can sum over $k_i$.  Surely, the result is degraded to an
asymptotic expansion only, but still we have an analytical solution
of the problem. This procedure was widely discussed in detail in
paper I \cite{paper1}, here we remind the result:

\begin {equation}
S_{N}=\sum\limits_{\mu=0}^{\mathcal{J}}\;\frac{(-1)^{\mu}}{\mu!\;(2z^2\triangle^3)^{\mu}}\;
\triangle^{3\mu}\left\{C^{2\mu}(N)\right\}_{2\mu,0} \label{re2}
\end {equation}
The evaluation of the symbols $\left\{C^{2\mu}(N)\right\}_{2\mu,0}$
is described in paper II \cite{paper2}. We explicitly present the
first nontrivial contribution in the continuum limit ($\mu$ finite,
$\triangle \rightarrow 0,\ \triangle N = \beta$):

\begin {equation}
\left\{C^2(a,b,c,\tau)\right\}_{2,0} = \frac{3}{8
\gamma^3}\left[3\gamma \tau \tanh^2(\gamma \tau) +
 \tanh(\gamma \tau) - \gamma \tau\right]
\end {equation}
where $\gamma=\sqrt{2b/c}.$ A typical $b$ dependence of the
$S(a,b,c,\tau)=\lim_{N\rightarrow\infty}\ S_N$ is expressed on the
Fig.1.

\begin{figure}
  \includegraphics[width=8cm]{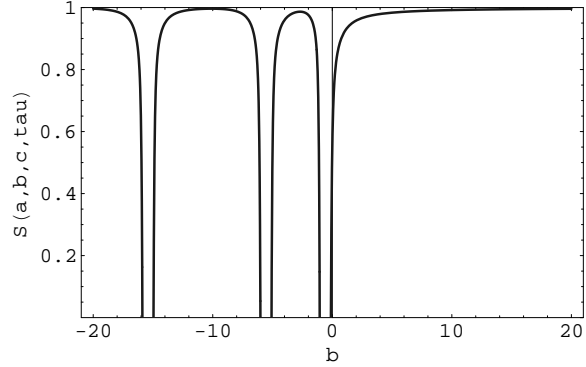}
  \caption {$b$ dependence of the continuum function $S(a,b,c,\tau)$ for fixed values $a, c, \tau.$ The first three nontrivial terms
 of the asymptotic series (\ref{re2}) were used.}
\end{figure}

 The parameters $a,c,\tau$ are fixed
constant, and we took $\mathcal{J}=3$ in (\ref{re2}). For $b<0$ we
see the singularities for $|\gamma\tau| = (n+1/2)\pi$. The divergent
behavior correspond to the powers of $\tan{(|\gamma\tau|)}$.

\subsection{Gel'fand - Yaglom Equation}

Gel'fand and Yaglom  proved  in \cite{gy} for the harmonic
oscillator that nontrivial continuum limit of the finite dimensional
integral approximation to the functional integral should be
evaluated from $N-$ dimensional integral results by a recurrent
procedure. For harmonic oscillator Gel'fand and Yaglom derived for
the unconditional measure integral $Z(\beta)$ the differential
equation:
\begin {equation}
\frac{\partial^2}{\partial \tau^2}y(\tau) = \frac{2b}{c}\  y(\tau),
\end {equation}
where $Z(\beta) = \lim_{N\rightarrow\infty}\ \mathcal{Z}_N =
y(\beta)^{-1/2}$

Following the idea of Gel'fand and Yaglom we found for an-harmonic
oscillator the generalized Gel'fand -- Yaglom equation. We define
the unconditional measure functional integral $Z(\beta)$ by
relation:
 $$Z(\beta)= \lim_{N\rightarrow\infty}\,
\mathcal{Z}_N = \frac{1}{\sqrt{F(\beta)}}\ ,\ F(\beta) =
\frac{y(\beta)}{S(\beta)^2}.$$

\noindent
The function $S(\tau)$ is given as the continuum limit of
Eq.(\ref{re2})
$$S(\tau)=\lim_{N \rightarrow \infty} \mathcal{S}_N\ .$$

\noindent
The generalized Gel'fand -- Yaglom equation read:
\begin {equation}
\frac{\partial^2}{\partial \tau^2}y(\tau) =
y(\tau)\left[\frac{2b}{c} + 4\left(\frac{\partial}{\partial
\tau}\ln{S(\tau)}\right)^2 \right]\ , \label{ggye}
\end {equation}
accompanied  by initial conditions $y(0) = S(0)^2,$ and $\left.
\left. \frac{\partial y(\tau)}{\partial \tau}\right|_{\tau=0} =
\frac{\partial}{\partial \tau}S(\tau)^2\right|_{\tau=0}. $

\noindent For $S(\tau)$ one can use a perturbative expansion in
coupling constant $a$ and then solve Eq. (\ref{ggye}). This
procedure gives a non-perturbative approximation of the functional
integral (\ref{fi}). For harmonic oscillator limit we have
$S(\tau)\rightarrow 1$.

 A typical dependence of the functions
$4\left(\frac{\partial}{\partial \tau}\ln{S(\tau)}\right)^2$ on
$\tau$ for positive fixed values of parameters $a, b, c$ is shown on
the figure 2.

\begin{figure}
  \includegraphics[width=8cm]{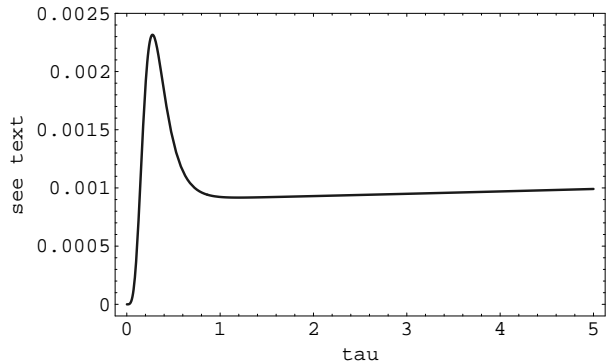}
 \caption{$\tau$ dependence of the continuum function
 $4[\partial_{\mu}\ln(S(a,b,c,\tau))]^2$ for fixed $a, b, c.$
 The first three nontrivial terms
 of the asymptotic series (\ref{re2}) were used.}
\end{figure}

\section{Conclusions}
We stressed some features  of the confinement problem in the
continuum theory, which can be solved by non-perturbative method of
evaluations of the functional integral. At the same time, in the
second part of the article we reported a new approach to the
solution of such problem.  We find for the functional integral of an
an-harmonic oscillator the non-perturbative equation (\ref{ggye}).
Solving this equation we in principle can find the analytical
solution of the an-harmonic oscillator problems as energy levels,
measurable quantities, etc, the studies are in progress.

 The skeptic view of
such solution of the problem may be expressed by the comment that we
reformulate the problem to the language of differential equations.
It is true. But our  opinion is that theory of the differential
equations is more elaborated and flexible than approaches based on
naive perturbative theory and it can give more reliable results than
the perturbative theory.

 \vskip 0.3cm {\bf{Acknowledgements}}. This work was
supported by VEGA projects No. 2/6074/26.

\end {document}